\documentclass[twocolumn,showpacs,superscriptaddress,preprintnumbers,amsmath,amssymb,prl,aps]{revtex4}

\usepackage{graphicx}
\usepackage{dcolumn}
\usepackage{bm}

\newcommand{\ipb}{\ensuremath{\mathrm{pb^{-1}}}}

\newcommand{\TeV}{\ensuremath{\matqq	hrm{Te\kern -0.1em V}}}
\newcommand{\TeVc}{\ensuremath{\mathrm{Te\kern -0.1em V\!/}c}}
\newcommand{\TeVcc}{\ensuremath{\mathrm{Te\kern -0.1em V\!/}c^2}}
\newcommand{\GeV}{\ensuremath{\mathrm{Ge\kern -0.1em V}}}
\newcommand{\GeVc}{\ensuremath{\mathrm{Ge\kern -0.1em V\!/}c}}
\newcommand{\GeVcc}{\ensuremath{\mathrm{Ge\kern -0.1em V\!/}c^2}}
\newcommand{\MeV}{\ensuremath{\mathrm{Me\kern -0.1em V}}}
\newcommand{\MeVc}{\ensuremath{\mathrm{Me\kern -0.1em V\!/}c}}
\newcommand{\MeVcc}{\ensuremath{\mathrm{Me\kern -0.1em V\!/}c^2}}

\newcommand{\cdfii}{CDF\,II~}

\newcommand{\ase}[2]{\ensuremath{^{~+ #1}_{~- #2}}}

\newcommand{\myto}{\kern -0.3em\to\kern -0.2em}
\newcommand{\Lxy}{L_{\rm xy}}
\newcommand{\BR}{{\mathrm {BR}}}

\newcommand{\phikpm}{\ensuremath{B^{\pm} \myto \phi K^{\pm}}}

\newcommand{\phik}{\ensuremath{B^+ \myto \phi K^+}}
\newcommand{\psik}{\ensuremath{{B^+ \myto J/\psi K^+}}}
\newcommand{\phiphi}{\ensuremath{B_s^0 \myto \phi \phi}}
\newcommand{\psiphi}{\ensuremath{B_s^0 \myto J/\psi \phi}}
\newcommand{\psikst}{\ensuremath{B^0 \myto J/\psi K^{\ast 0}}}
\newcommand{\phikst}{\ensuremath{B^0 \myto \phi K^{\ast 0}}}
\newcommand{\kstar}{\ensuremath{K^{\ast 0} }}

\newcommand{\mpmm}{\ensuremath{\mu^+\mu^-}}
\newcommand{\KKpm}{\ensuremath{K^+K^-}}

\newcommand{\ACP}{\ensuremath{A_{CP}}}
\newcommand{\DGs}{\ensuremath{\Delta \Gamma_s}}

\newcommand{\BsVV}{\ensuremath{B_s^0\rightarrow VV\ }}

\newcommand{\phiKK}{\ensuremath{\phi \myto \KKpm}}
\newcommand{\psimm}{\ensuremath{J/\psi \myto \mpmm}}

\begin{document}

\title{
{\raggedleft\small
{\tt FERMILAB-PUB-05-027-E\\
CDF/PUB/BOTTOM/PUBLIC/7395}\\
}\vspace*{0.9cm}
First Evidence for \phiphi\ Decay and Measurements of Branching Ratio and 
\ACP\ for \phik} 


\affiliation{Institute of Physics, Academia Sinica, Taipei, Taiwan 11529, Republic of China }
\affiliation{Argonne National Laboratory, Argonne, Illinois 60439 }
\affiliation{Institut de Fisica d'Altes Energies, Universitat Autonoma de Barcelona, E-08193, Bellaterra (Barcelona), Spain }
\affiliation{Istituto Nazionale di Fisica Nucleare, University of Bologna, I-40127 Bologna, Italy }
\affiliation{Brandeis University, Waltham, Massachusetts 02254 }
\affiliation{University of California at Davis, Davis, California 95616 }
\affiliation{University of California at Los Angeles, Los Angeles, California 90024 }
\affiliation{University of California at San Diego, La Jolla, California 92093 }
\affiliation{University of California at Santa Barbara, Santa Barbara, California 93106 }
\affiliation{Instituto de Fisica de Cantabria, CSIC-University of Cantabria, 39005 Santander, Spain }
\affiliation{Carnegie Mellon University, Pittsburgh, PA 15213 }
\affiliation{Enrico Fermi Institute, University of Chicago, Chicago, Illinois 60637 }
\affiliation{Joint Institute for Nuclear Research, RU-141980 Dubna, Russia }
\affiliation{Duke University, Durham, North Carolina 27708 }
\affiliation{Fermi National Accelerator Laboratory, Batavia, Illinois 60510 }
\affiliation{University of Florida, Gainesville, Florida 32611 }
\affiliation{Laboratori Nazionali di Frascati, Istituto Nazionale di Fisica Nucleare, I-00044 Frascati, Italy }
\affiliation{University of Geneva, CH-1211 Geneva 4, Switzerland }
\affiliation{Glasgow University, Glasgow G12 8QQ, United Kingdom }
\affiliation{Harvard University, Cambridge, Massachusetts 02138 }
\affiliation{Division of High Energy Physics, Department of Physics, University of Helsinki and Helsinki Institute of Physics, FIN-00044, Helsinki, Finland }
\affiliation{Hiroshima University, Higashi-Hiroshima 724, Japan }
\affiliation{University of Illinois, Urbana, Illinois 61801 }
\affiliation{The Johns Hopkins University, Baltimore, Maryland 21218 }
\affiliation{Institut f\"ur Experimentelle Kernphysik, Universit\"at Karlsruhe, 76128 Karlsruhe, Germany }
\affiliation{High Energy Accelerator Research Organization (KEK), Tsukuba, Ibaraki 305, Japan }
\affiliation{Center for High Energy Physics: Kyungpook National University, Taegu 702-701; Seoul National University, Seoul 151-742; and SungKyunKwan University, Suwon 440-746; Korea }
\affiliation{Ernest Orlando Lawrence Berkeley National Laboratory, Berkeley, California 94720 }
\affiliation{University of Liverpool, Liverpool L69 7ZE, United Kingdom }
\affiliation{University College London, London WC1E 6BT, United Kingdom }
\affiliation{Massachusetts Institute of Technology, Cambridge, Massachusetts 02139 }
\affiliation{Institute of Particle Physics: McGill University, Montr\'eal, Canada H3A~2T8; and University of Toronto, Toronto, Canada M5S~1A7 }
\affiliation{University of Michigan, Ann Arbor, Michigan 48109 }
\affiliation{Michigan State University, East Lansing, Michigan 48824 }
\affiliation{Institution for Theoretical and Experimental Physics, ITEP, Moscow 117259, Russia }
\affiliation{University of New Mexico, Albuquerque, New Mexico 87131 }
\affiliation{Northwestern University, Evanston, Illinois 60208 }
\affiliation{The Ohio State University, Columbus, Ohio 43210 }
\affiliation{Okayama University, Okayama 700-8530, Japan }
\affiliation{Osaka City University, Osaka 588, Japan }
\affiliation{University of Oxford, Oxford OX1 3RH, United Kingdom }
\affiliation{University of Padova, Istituto Nazionale di Fisica Nucleare, Sezione di Padova-Trento, I-35131 Padova, Italy }
\affiliation{University of Pennsylvania, Philadelphia, Pennsylvania 19104 }
\affiliation{Istituto Nazionale di Fisica Nucleare Pisa, Universities of Pisa, Siena and Scuola Normale Superiore, I-56127 Pisa, Italy }
\affiliation{University of Pittsburgh, Pittsburgh, Pennsylvania 15260 }
\affiliation{Purdue University, West Lafayette, Indiana 47907 }
\affiliation{University of Rochester, Rochester, New York 14627 }
\affiliation{The Rockefeller University, New York, New York 10021 }
\affiliation{Istituto Nazionale di Fisica Nucleare, Sezione di Roma 1, University of Roma ``La Sapienza," I-00185 Roma, Italy }
\affiliation{Rutgers University, Piscataway, New Jersey 08855 }
\affiliation{Texas A\&M University, College Station, Texas 77843 }
\affiliation{Texas Tech University, Lubbock, Texas 79409 }
\affiliation{Istituto Nazionale di Fisica Nucleare, University of Trieste/\ Udine, Italy }
\affiliation{University of Tsukuba, Tsukuba, Ibaraki 305, Japan }
\affiliation{Tufts University, Medford, Massachusetts 02155 }
\affiliation{Waseda University, Tokyo 169, Japan }
\affiliation{Wayne State University, Detroit, Michigan 48201 }
\affiliation{University of Wisconsin, Madison, Wisconsin 53706 }
\affiliation{Yale University, New Haven, Connecticut 06520 }


\author{D.~Acosta}
\affiliation{University of Florida, Gainesville, Florida 32611 }

\author{J.~Adelman}
\affiliation{Enrico Fermi Institute, University of Chicago, Chicago, Illinois 60637 }

\author{T.~Affolder}
\affiliation{University of California at Santa Barbara, Santa Barbara, California 93106 }

\author{T.~Akimoto}
\affiliation{University of Tsukuba, Tsukuba, Ibaraki 305, Japan }

\author{M.G.~Albrow}
\affiliation{Fermi National Accelerator Laboratory, Batavia, Illinois 60510 }

\author{D.~Ambrose}
\affiliation{Fermi National Accelerator Laboratory, Batavia, Illinois 60510 }

\author{S.~Amerio}
\affiliation{University of Padova, Istituto Nazionale di Fisica Nucleare, Sezione di Padova-Trento, I-35131 Padova, Italy }

\author{D.~Amidei}
\affiliation{University of Michigan, Ann Arbor, Michigan 48109 }

\author{A.~Anastassov}
\affiliation{Rutgers University, Piscataway, New Jersey 08855 }

\author{K.~Anikeev}
\affiliation{Fermi National Accelerator Laboratory, Batavia, Illinois 60510 }

% ...

\collaboration{CDF Collaboration}

\date{May 24, 2005}

\begin{abstract}
We present the first evidence of charmless decays of 
the $B_s^0$ meson,  the decay \phiphi, and a 
measurement of the Branching Ratio $\BR(\phiphi)$ 
using $180\,{\rm pb}^{-1}$ of data collected 
by the \cdfii experiment at the Fermilab Tevatron collider. 
In addition, the BR and direct $CP$~asymmetry for the \phik\ decay
are measured. We obtain: 
$\BR (\phiphi) = (14 \ase{6}{5}(stat.) \pm 6(syst.)) \times 10^{-6}$,
$\BR (\phik) = (7.6 \pm 1.3(stat.) \pm 0.6(syst.) ) \times 10^{-6}$,
and $\ACP(\phik) = -0.07 \pm 0.17 (stat.) \ase{0.03}{0.02} (syst.)$.
Both decays are governed in the Standard Model by second order (penguin) 
$b \myto s\bar{s}s$ amplitudes.
\end{abstract}


\maketitle

%
Recent measurements~\cite{Belle-btos-ICHEP04,Babar-btos-ICHEP04} 
of $CP$ asymmetries for
$B^0 \myto \phi K^0_S$ and $B^0 \myto \eta^\prime K^0_S$
suggest deviations from Standard Model (SM) predictions in
$B$~meson decays involving $b \myto s\bar{s}s$ transitions,
which also generate both \phiphi\ and \phik\
decays.
Charmless decays of $B^0$ and $B^+$ mesons provide
among the most stringent tests of the CKM matrix. 
Charmless $B_s^0$ decays, not yet experimentally 
established, offer important additional 
ways of testing the theory. Of particular interest for the 
current and next generation experiments at hadron colliders
are charmless $B_s^0$ to Vector-Vector decays into 
self-conjugate final states, like \phiphi.
Exploiting angular correlations between the final state particles
in \BsVV decays allows the statistical separation of the 
$CP$-even and $CP$-odd components of the decay amplitude.
With sufficient statistics, this decompostion allows the 
measurement of the $B_s^0$ decay width difference (\DGs), CKM studies, and
tests of decay polarization predictions~\cite{BVV-theory}.
In particular, the \phiphi\ decay, which proceeds through a pure 
$b \myto s\bar{s}s$, has been considered as a probe for New Physics 
scenarios ~\cite{phiphi-Theory} 
(e.g. in $b \myto s$ penguin dominated decays) recently
advocated to explain the observed $b \myto s\bar{s}s$ asymmetries.
%
The measurement of the branching ratio $\BR(\phiphi)$ gives insight into the
size of penguin amplitudes as well as tightens the constraints on 
poorly known form factors.
Recent calculations 
predict $\BR(\phiphi)$ between 
$10\times 10^{-6}$~\cite{phiphi-PQCD} and $37\times 10^{-6}$~\cite{phiphi-BR}.
Within the SM the \phik\ direct $CP$ asymmetry ($\ACP$) is predicted 
not to exceed a few percent~\cite{phiktheory}. 
No single experiment~\cite{pdg2004} has
yet reached the sensitivity to detect this asymmetry. Further 
measurements of decay rates and \ACP\ may help identifying
the origin of the observed deviations from 
SM predictions in $b \myto s\bar{s}s$ penguin dominated 
$B$ decays.
\newline
%
%
\indent We report the first evidence of \phiphi\ decays, and present
the first measurements of the $CP$-averaged BR and \ACP\ for
\phik~\cite{charge-conjugate-disclaimer} at hadron colliders.
To cancel the uncertainty in the $B$~hadron production cross section and
to reduce systematic uncertainties on detector efficiencies,
the branching fractions are extracted from ratios of the decay
rates of interest normalized to the established
\psiphi\ and \psik\ decay modes, which are characterized by the
same number of decay vertices and charged tracks in the final state
as in the signal modes.
\newline
\indent 
In this analysis we use 180~\ipb\ of $p\bar p$~collision data
at $\sqrt{s}=1.96$~TeV collected by the upgraded 
Collider Detector (CDF\,II) at the Fermilab Tevatron.
The components of the \cdfii detector pertinent to this analysis are 
briefly described. A more complete description can be found 
elsewhere~\cite{cdf}. 
We use  tracks in the pseudorapidity 
range $|\eta|\lesssim 1$~\cite{CDFsystem} reconstructed by a silicon microstrip 
vertex detector (SVX\,II)~\cite{svxii} 
and the Central Outer Tracker (COT)~\cite{cot}, 
which are immersed in a $1.4\,{\rm T}$ solenoidal magnetic field. 
The SVX\,II detector consists of double-sided  
sensors arranged in five cylindrical layers.
Surrounding the SVX\,II  is the COT, an open cell drift chamber with 
$96$ sense wires.
The integrated charge collected by each wire provides a 
measurement of the specific ionization ($dE/dx$) for charged particles, 
allowing a separation equivalent to 
1.4 Gaussian $\sigma$ between $\pi$ and $K$ for $p_T > 2\ \GeVc$.
%
%
A set of planar drift chambers, located outside the calorimeters and 
additional steel absorbers, is used to detect muons within
$|\eta|\le 1$ with high purity. \newline
%
%
\indent 
A sample enriched with heavy flavor particles is 
selected by the three-level displaced track trigger.
At Level~1, charged tracks are reconstructed in the COT 
by the eXtremely Fast Tracker (XFT)~\cite{xft}.
The trigger requires two oppositely charged tracks with transverse
momenta $p_T\ge2\,\GeVc$ and the scalar sum 
$p_{T1}+p_{T2}\ge5.5\,\GeVc$. At 
Level~2, the Silicon Vertex Tracker (SVT)~\cite{svt} associates SVX\,II
$r$-$\phi$ position measurements with XFT tracks, providing 
a precise measurement of the track impact parameter ($d_0$), 
the distance of closest approach of the track trajectory to 
the beam axis in the transverse plane.
Decays of heavy flavor particles are identified by requiring 
two tracks with $120\,\mu{\rm m}\le d_0 \le1.0\rm\,mm$ and 
an opening angle $2^\circ\le|\Delta\phi|\le90^\circ$.
A requirement $\Lxy >200\rm\,\mu m$ is also applied,
where the two-dimensional decay length, $\Lxy$, is calculated
as the transverse distance from the beam axis to the two track 
intersection projected onto the total transverse momentum of the 
track pair.
A complete event reconstruction is performed at Level 3, where the
Level 1 and Level 2 trigger requirements are confirmed.
%
%
\newline
\indent $B$~candidates are reconstructed by detecting \phiKK\ and \psimm\
decays. In the latter case at least one of the muons has to be identified 
in the muon detectors 
to suppress contamination from other two-body $J/\psi$ decays. 
At least one pair of tracks ("trigger tracks"), has to satisfy 
the trigger requirements. $B^+$ ($B_s^0$) candidates are 
formed by fitting three (four) tracks with $p_T > 0.4\, \GeVc$ to 
a common vertex. 
%
Requiring a good vertex fit $\chi^2$ reduces background from mismeasured tracks.
Combinatoric background is reduced by exploiting several
variables sensitive to the long lifetime and relatively
hard $p_T$ spectrum of $B$ mesons and the isolation of $B$~hadrons inside 
$b$-quark jets. For this purpose we define the quantity $I_{R}$ as the
ratio of the $B^+$ candidate $p_{T}$ over the total transverse momenta 
of all tracks within a cone of radius $R=\sqrt{\Delta\eta^2+\Delta\phi^2}=1$ 
around the $B$~flight direction. 
Requiring the $B$~flight direction to extrapolate back to the beam axis
decreases background from partially reconstructed decays. 
The cut values on the discriminating variables are optimized by maximizing 
$\mathcal {S/\sqrt{S+B}}$ for the already observed \phik\ signal 
and $\mathcal{S}/(1.5+\sqrt{\mathcal{B}})$ for \phiphi\ 
whose branching ratio is unknown. 
The latter choice is equivalent to maximizing the  
potential to reach a $3\sigma$ observation of a new signal~\cite{Punzi}.
The signal ($\mathcal{S}$) is derived from a
Monte Carlo (MC) simulation~\cite{cdfsim} of the \cdfii detector and trigger
that uses the $B$~meson momentum and rapidity distributions from Ref.~\cite{MNR},
which were matched to CDF data.
The background ($\mathcal{B}$) is represented by appropriately normalized 
data selected with the same requirements as the signal except 
for the two-kaon invariant mass lying in the 
$\phi$ sideband region: $1.04 < m_{KK} < 1.06\ \GeVcc$. 
\newline
\begin{figure}
\includegraphics[width=1.00\linewidth]{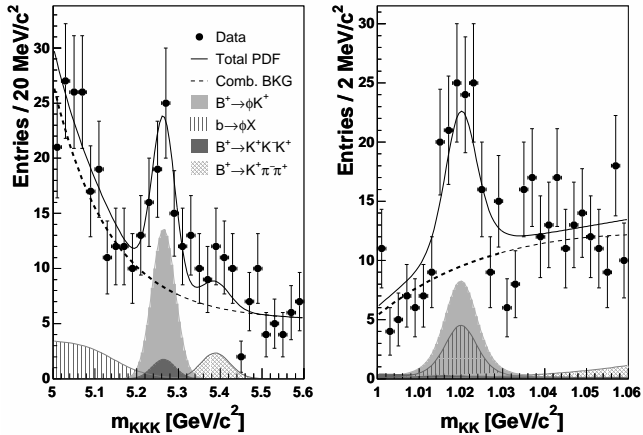} 
\caption{\label{fig_phik}
$m_{KKK}$ (left) and $m_{KK}$ (right) distributions 
for \phik\ with projections of the total likelihood function and its 
components. $B^+\myto K^+K^-K^+$ is the sum of 
non-resonant and $B^+\myto f_0(980) K^+$ contributions, while
$B^+\myto K^+\pi^-\pi^+$ includes non-resonant and
$B^+\myto K^{\ast 0}(892) \pi^+$.
}
\end{figure}
\indent
 We discuss first the \phik\ analysis.
The optimization results in the following requirements:  
vertex $\chi^2 < 8$, $B^+$~decay 
length $\Lxy > 350$ $\mu$m, $B^+$ reconstructed impact 
parameter $d_0^B < 100$ $\mu$m, isolation  $I_{R} > 0.5$,
non-trigger track transverse momentum $p_T^{\text{\it soft}}> 1.3$ $\GeVc$ and
impact parameter $d_0^{\text{\it soft}}> 120$ $\mu$m. 
\newline
\indent
The \phikpm\ total yield and \ACP, defined as
\begin{eqnarray*}
\label{eq:ACPphik}
\ACP \equiv \frac{N(B^-\to\phi K^-) - N(B^+\to\phi K^+)}
{N( B^-\to\phi K^-) + N(B^+\to\phi K^+)} \ ,
\end{eqnarray*}
are extracted simultaneously from an extended unbinned maximum likelihood fit
on four variables to the combined $B^+$ and $B^-$ sample: 
the three-kaon invariant mass ($m_{KKK}$), the invariant mass of the
$\phi$ candidate ($m_{KK}$), the cosine of the $\phi$ meson helicity angle
(defined as the angle between the $K^+$ momentum in the parent $\phi$
rest frame and the momentum of the $\phi$ in the $B$ rest frame)
and the measured $dE/dx$ deviation from the expected value for pions 
for the lowest momentum trigger track.
The likelihood function has seven components: signal, partially reconstructed 
$b\myto \phi X$ decays, combinatoric background, 
$B^+\myto K^{\ast 0}(892) \pi^+$, with $\kstar \myto K^+\pi^-$, 
$B^+\myto f_0(980) K^+$, with $f_0\myto K^+K^-$,  
non-resonant $B^+\myto K^+K^-K^+$ and non-resonant $B^+\myto K^+\pi^-\pi^+$.
The normalizations of the last three components are fixed to the
$B^+\myto K^{\ast 0}(892) \pi^+$ yield, determined in the fit, through
their relative decay rates and detection efficiencies. 
For each component the likelihood function 
is the product of four one-dimensional probability density functions (PDF)
of the fit variables, which are assumed to be uncorrelated.
The $m_{KKK}$ and $m_{KK}$ distributions are shown in Fig.~\ref{fig_phik}
with projections for the different components.
\newline
\indent A combination of Monte Carlo simulation and sideband
data is used to derive the PDF
in each variable for the various fit components.
For $m_{KKK}$ the fully reconstructed signals 
are modeled by Gaussian functions. The mass and width of the $B$~signals 
are determined from the fit in the case of $KKK$ final states,
while for $K\pi\pi$ they are fixed to the values predicted by the simulation.
We derive a parameterization from simulation for the partially reconstructed 
decays that populate the low mass side of the $m_{KKK}$ distribution. 
An exponential plus a constant describe the combinatoric background.
In the PDF for $m_{KK}$, the $\phi$ resonance is described by a Breit-Wigner
convoluted with a Gaussian resolution function, while the combinatoric
background is modeled by an empirical phase space function. 
Shapes for other backgrounds are derived from simulation.
The $\phi$ helicity PDFs for $B$ decays are derived from simulation, while 
the combinatoric background PDF is modeled using data from the $\phi$ sideband. 
The $dE/dx$ PDFs for kaons and pions are derived from 
a high statistics $D^0 \myto K^-\pi^+$ sample 
obtained from $D^{\ast\pm}$ decays.
We find $N_{\phi K}=47.0\pm 8.4$, $\ACP = -0.07 \pm 0.17$ and
$N_{K^{\ast 0}\pi^+} = 7.8 \pm 6.0$ (statistical errors only) from which we
estimate a $B^+\myto f_0(980) K^+$ contamination of 11\% under the signal peak.
\newline
\indent 
Candidates for the normalization mode, \psik, are selected with 
the same requirements as the \phik\ candidates except for the invariant 
mass of the two muons being within 100 \MeVcc\ of the $J/\psi$ mass~\cite{pdg2004}.
Using an extended likelihood fit of the $m_{\mu\mu K}$ and $m_{\mu\mu}$ 
distributions, we obtain a total yield of $N_{\psi K}=439\pm 22$
(statistical error only). The asymmetry $\ACP(\psik) = 0.046 \pm 0.050$ 
is consistent with zero as expected for this mode.
\newline
\indent
The relative \phik\ decay rate is calculated using:
\begin{eqnarray*}
\label{eq:BRphik}
\frac{\BR\left(\phik \right)}{\BR\left(\psik \right)} =
\frac{N_{\phi K}}{N_{\psi K}}
\frac{\BR\left(J/\psi\to\mu\mu\right)}{\BR \left(\phi\to K K \right)}
\frac{\epsilon_{\psi K}}{\epsilon_{\phi K}} \, \epsilon^\mu_{\psi K},
\end{eqnarray*}
where $\epsilon_{\psi K}/\epsilon_{\phi K}=0.721\pm 0.011$ is the ratio of 
the combined trigger and selection efficiencies derived from MC with 
a correction of about 5\% due to the different 
trigger efficiency for muons and kaons as measured in unbiased samples.
The efficiency for identifying at least one of the decay muons,
$\epsilon^\mu_{\psi K}=0.81\pm 0.02$, is obtained by weighting the expected 
$p_T$ spectra in our signal with the single muon identification efficiency
measured as a function of $p_T$ in a sample of inclusive \psimm\ decays. 
The relative $\BR(\phik)$ and $\ACP$ results are 
reported in Table~\ref{tab:summResults}.
%
%
\begin{figure*} 
\includegraphics[width=0.42\linewidth]{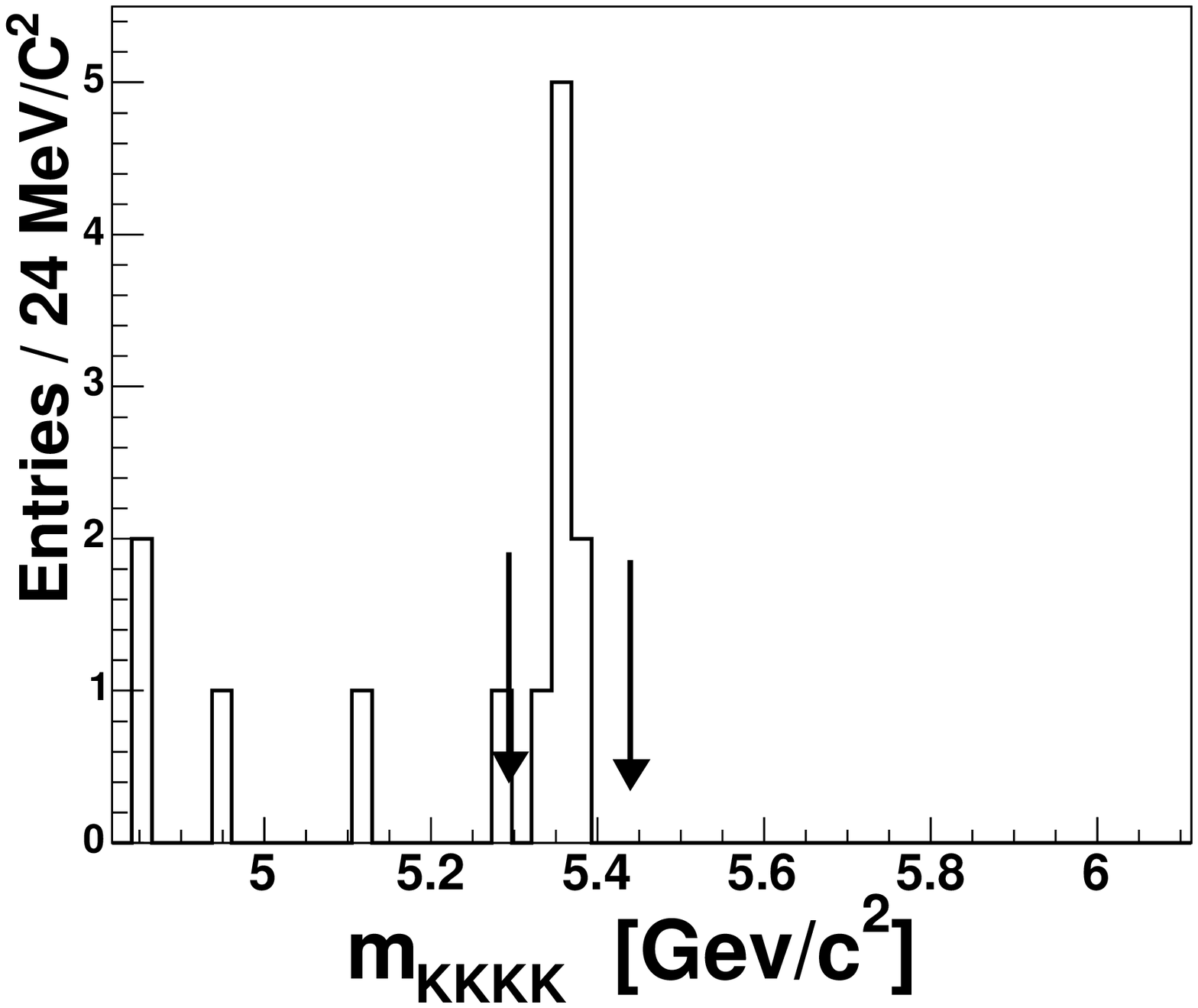} 
\includegraphics[width=0.42\linewidth]{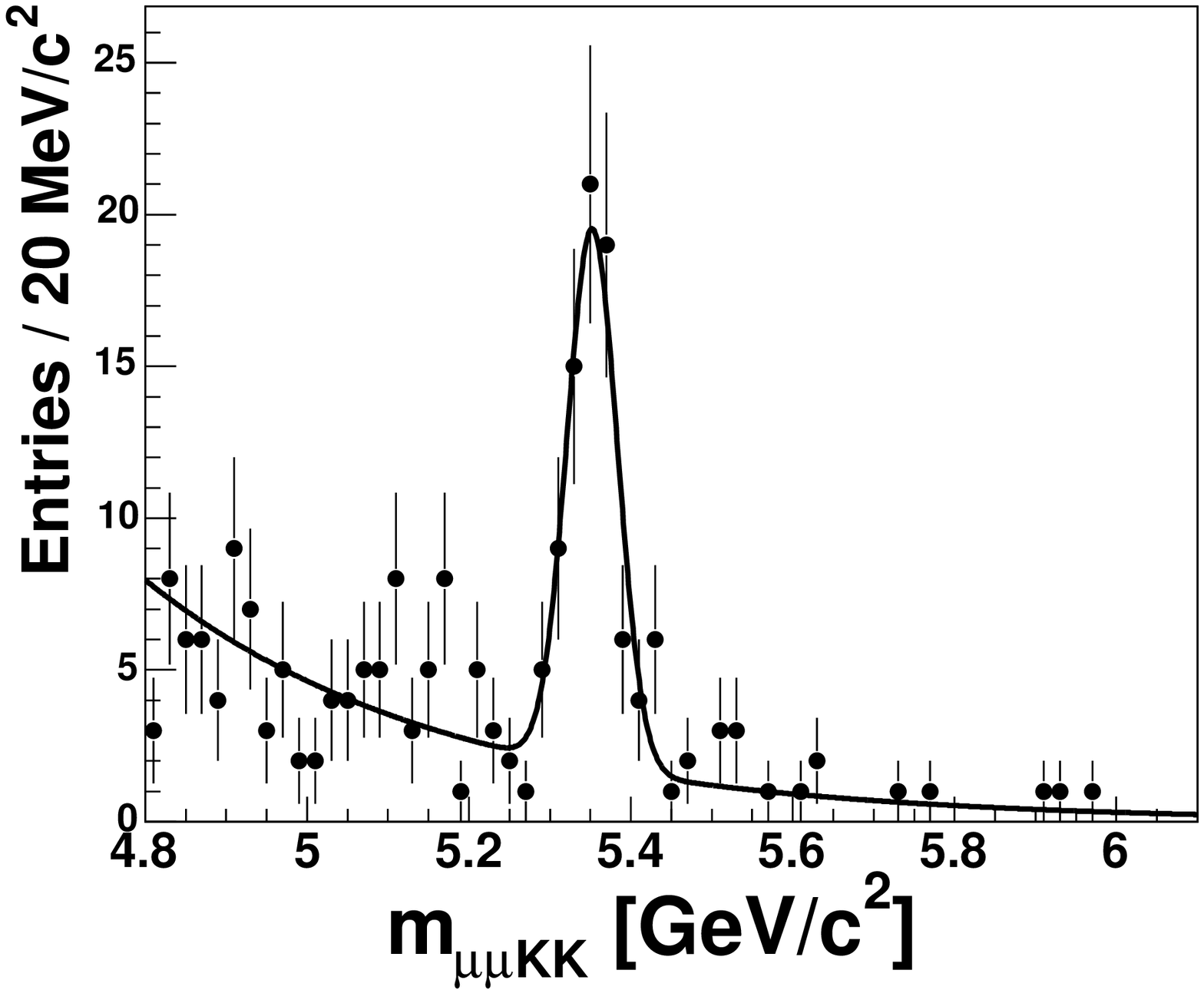} 
\caption{\label{fig_phiphi}
Invariant mass distribution for $\phiphi$ (left) and $\psiphi$ (right)
candidates. The arrows indicate the signal region for the 
$\phiphi$ search. The fit described in the text is superimposed 
on the $m_{\mu\mu K K}$ distribution.}
\end{figure*}
Systematic uncertainties on signal yield ($\pm 1.4$ events) and asymmetry 
(+0.03, -0.02) are evaluated by varying the PDFs used in the likelihood fit, 
including a variation of the $f_0(980)$ width from 40 to 
100~\MeVcc. The uncertainties on the determination of efficiency 
and muon identification introduce a 5.6\% relative error in the 
measurement of the branching ratio and have no effect for \ACP. 
The uncertainties from BR($\phi \myto KK$) and BR($J/\psi \myto \mu\mu$) 
contribute an additional 2\% to the total systematic.
%
%
We determine a $\pm 0.005$ uncertainty on the \ACP\ measurement 
arising from a possible tracking efficiency asymmetry
for $K^+$ and $K^-$ as measured in minimum bias data.
\begin{table}[h]
\caption{\label{tab:summResults} 
Results for \phik\ and \phiphi\ decays. World average~\cite{pdg2004} 
$\phi\myto K^+K^-$ and $J/\psi\myto \mpmm$ BR have been used to obtain 
the BR ratio $\BR(\phik)/\BR(\psik)$ and $\BR(\phiphi)/\BR(\psiphi)$. 
The first uncertainty is statistical, the second is systematic.}
\begin{ruledtabular}
\begin{tabular}{lcc}
                          & $ \phik  $  &  $\phiphi $ \\\hline
Yield                     & $47.0 \pm 8.4 \pm 1.4$                
                          & $ 7.3 \ase {3.2}{2.5} \pm 0.4 $ \\
$\BR\, {\rm ratio}$       & $ (7.6 \pm 1.3 \pm 0.6)\times 10^{-3} $ 
                          & $ (10 \ase{5}{4} \pm 1)\times 10^{-3}$ \\
$\ACP$                    & $-0.07 \pm 0.17 ^{+0.03}_{-0.02}$ 
                          &  --  \\
\end{tabular}
\end{ruledtabular}
\end{table}
\newline
\indent
A ``blind'' search for \phiphi\ decays was performed fixing the selection 
requirements and evaluating the combinatoric background from 
independent samples before examining the signal region in the data.
The signal is selected requiring two pairs of kaons 
having an invariant mass within 15 \MeVcc\ of the world average $\phi$ 
mass~\cite{pdg2004}. 
The optimized selection criteria are the following:
vertex 
$\chi^2 < 10$, $B_s^0$~decay length
$\Lxy > 350\,\mu$m, $B_s^0$~reconstructed impact parameter $d_0^B<80\,\mu$m,
minimum $\phi$ meson transverse momentum $p_T^{\phi} >2.5\,\GeVc$ and 
minimum of the two kaons' impact parameters from each $\phi$ meson candidate
$d_{0\,min}^{\phi_1}> 40\,\mu$m, $d_{0\,min}^{\phi_2}> 110\,\mu$m,
where $\phi_1$ is the lower momentum $\phi$ candidate.
The \phiphi\  candidate mass distribution is shown in 
Fig.~\ref{fig_phiphi}.
In a region of $\pm 72\, \MeVcc$ around the world average~\cite{pdg2004}
$B_s^0$ mass,
corresponding to a window three times the 
expected mass resolution, we observe 8 events. \newline
\indent Two sources of background are expected in the  $B_s^0$~signal region: 
combinatoric background 
and \phikst\ decays with the pion from the \kstar\ decay 
treated as a kaon.
The combinatoric background generates
a smooth distribution in the invariant mass region close
to $m_{B_s^0}$. Its contribution in the signal region is 
$0.35 \pm 0.37$ events (combining statistical and systematic
uncertainties), estimated using a background enriched sample where 
both $\phi$ meson candidates have invariant mass
lying in the $\phi$ mass sideband region.
The \phikst\ background results in an approximately Gaussian distribution 
underneath the $B_s^0$ signal. Its contribution, derived from 
simulation, is $0.37 \pm 0.18$ events where the error includes 
both statistical and systematic uncertainties,
resulting in a signal yield of $7.3 \ase {3.2}{2.5}$ events.
The probability of a Poisson fluctuation of the background 
to the observed or higher number of events is $1.3 \times 10^{-6}$, 
corresponding to 4.7 $\sigma$ one-sided Gaussian significance. 
%
%
\newline
\indent For the determination of $\BR(\phiphi)$, 
a normalization sample of \psiphi\ decays is selected requiring one pair of kaons
and one pair of muons within 15 and 50 \MeVcc\ of the world average $\phi$ and
$J/\psi$ mass respectively. The other kinematic selection criteria 
are similar to the \phiphi\ mode.
To extract the number of \psiphi\ events, the candidate invariant mass 
distribution, shown in Fig.~\ref{fig_phiphi}, is fit with a binned maximum 
likelihood function using a Gaussian for the signal and an exponential 
for the background.
The fit returns 
$N_{\psi\phi} = 69 \pm 10(stat.) \pm 5 (syst.)$ events, where the 
systematic error is evaluated using alternative background models
for the low mass region where partially reconstructed 
$B$~decays are expected.  
We subtract $3.7\pm1.7$  background events from \psikst\ decays, 
with the pion treated as a kaon, as estimated from simulation.
\newline
\indent The \phiphi\ decay rate is derived from the relation:
\begin{eqnarray*}
\label{eq:BRphiphi}
\frac{\BR\left(\phiphi\right)}
     {\BR\left(\psiphi\right)} =
\frac{N_{\phi\phi}}{N_{\psi\phi}}
\frac{\BR\left(J/\psi\to\mu\mu\right)}{\BR \left(\phi\to K K \right)}
\frac{\epsilon_{\psi\phi}}{\epsilon_{\phi\phi}}\, \epsilon^{\mu}_{\psi\phi}\ ,
\end{eqnarray*}
%
%
where $\epsilon_{\phi\phi}/\epsilon_{\psi\phi}=0.821\pm 0.015$ is 
derived from simulation as in the \phik\ case and 
$\epsilon^{\mu}_{\psi\phi} = 0.92 \pm 0.05$ is obtained
using the $p_T$ spectra of the \psiphi\ signal and the 
muon identification efficiency curve discussed above. 
\newline
\indent The uncertainty on the \psiphi\ yield and  
background evaluation contribute 8\% to the relative systematic 
error. The efficiencies significantly depend on
both the polarization of the decay vector particles and the assumed 
value of \DGs\ since they affect the impact parameter 
of the decay products on which the trigger operates.
We vary the longitudinal polarization of the 
\phiphi\ decay from 20\% to 80\%, the \psiphi\ polarization amplitudes 
within $1\sigma$ of the CDF measurement~\cite{CDF-psiphiPol} and
\DGs\ in the range $0.06<\DGs/\Gamma_s<0.18$ to assign
a relative systematic uncertainty of 4\%.
Summing in quadrature all contributions, 
we estimate a total relative systematic uncertainty on
the ratio of BR(\phiphi) to BR(\psiphi) of 11\%. 
To convert the ratio BR(\phiphi)/BR(\psiphi) reported in 
Table~\ref{tab:summResults} in a measurement of BR(\phiphi),  
we use $\BR(\psiphi)=(1.38 \pm 0.49) \times 10^{-3}$, obtained from
correcting the CDF measurement~\cite{CDF-psiphiBR} for the current
value of $f_s/f_d$, the $B_s^0$ to $B^0$ production ratio~\cite{pdg2004}.
We derive:
$\BR (\phiphi) = ( 14 \ase{6}{5}(stat.) \pm 6(syst.) ) \times 10^{-6}$, 
where the systematic uncertainty includes a 36\% contribution 
due to the uncertainty on BR(\psiphi). 
\newline
\indent 
In summary, we have used $p\bar p$ collision data collected with the
displaced track trigger of the \cdfii detector to study 
two fully reconstructed $b \myto s\bar{s}s$ penguin dominated $B$ decays. 
Using the world average~\cite{pdg2004} BR(\psik) we measure:
$\BR (\phik) = (7.6 \pm 1.3(stat.) \pm 0.6(syst.) ) \times 10^{-6}$ and
$\ACP(\phik) = -0.07 \pm 0.17 (stat.) \ase{0.03}{0.02} (syst.)$
which agree with previous measurements~\cite{pdg2004}.
We find the first evidence of a charmless Vector-Vector $B_s^0$ decay and 
measure BR(\phiphi) in agreement with the estimate of Ref.~\cite{phiphi-PQCD}
and the recently amended calculation in~\cite{phiphi-BR}. 
\newline
\indent 
We thank the Fermilab staff and the technical staffs of the 
participating institutions for their vital contributions. 
This work was supported by the U.S. Department of Energy and 
National Science Foundation; the Italian Istituto Nazionale di 
Fisica Nucleare; the Ministry of Education, Culture, Sports, 
Science and Technology of Japan; the Natural Sciences and 
Engineering Research Council of Canada; the National Science 
Council of the Republic of China; the Swiss National Science Foundation; 
the A.P. Sloan Foundation; 
the Bundesministerium f\"ur Bildung und Forschung, Germany; 
the Korean Science and Engineering Foundation and 
the Korean Research Foundation; 
the Particle Physics and Astronomy Research Council and the Royal Society, UK;
the Russian Foundation for Basic Research; 
the Comision Interministerial de Ciencia y Tecnologia, Spain; 
and in part by the European Community's Human Potential Programme 
under contract HPRN-CT-2002-00292, Probe for New Physics.


\begin{thebibliography}{99}


\bibitem{Belle-btos-ICHEP04}
K.~Abe {\it et al.}  [BELLE Collaboration],
arXiv:hep-ex/0409049.

\bibitem{Babar-btos-ICHEP04} 
B.~Aubert {\it et al.}  [BABAR Collaboration],
arXiv:hep-ex/0408072 and arXiv:hep-ex/0408090.

\bibitem{BVV-theory} 
A.\,S.\, Dighe, I.\, Dunietz and R.\, Fleisher, Eur. Phys. J.
{\bf C6}, 647 (1999);
D.\,Atwood and A.\,Soni,
Phys.\ Rev.\ D {\bf 65}, 073018 (2002).

\bibitem{phiphi-Theory}
M.\, Raidal,
Phys.\ Rev.\ Lett.\  {\bf 89}, 231803 (2002);
A.\, Datta {\it et al.}, arXiv:hep-ph/0406192; submitted to Phys.\ Rev.\ D.

\bibitem{phiphi-PQCD}
Y.-H.\, Chen {\it et al.}, Phys. Rev. D {\bf 59}, 074003 (1999).

\bibitem{phiphi-BR}
X.~Q.~Li, G.~R.~Lu and Y.~D.~Yang,
Phys.\ Rev.\ D {\bf 68}, 114015 (2003)
[Erratum-ibid.\ D {\bf 71}, 019902 (2005)].


\bibitem{phiktheory} 
M.\,Beneke and M.\,Neubert,
Nucl.\ Phys.\ B {\bf 675}, 333 (2003).
%
%


\bibitem{pdg2004} 
S.\,Eidelman {\it et al.} [Particle Data Group],
Phys.\ Lett.\ B {\bf 592}, 1 (2004).

\bibitem{charge-conjugate-disclaimer} Charge conjugate decay modes are 
implied throughout this letter unless otherwise stated.

\bibitem{cdf}
D. Acosta et al., Phys. Rev. D71, 032001 (2005).


\bibitem{CDFsystem}
\cdfii uses a cylindrical coordinate system in which $\phi$ is the azimuthal
angle, $r$ is the radius from the nominal beam axis, $y$ points up and $z$
points in the proton beam direction with the origin at the center of the 
detector.The transverse plane is the plane perpendicular to the $z$-axis.

\bibitem{svxii}
A.\,Sill {\it et al.}, Nucl. Instrum. Meth. A {\bf 447}, 1 (2000).
                                                                  
\bibitem{cot}
T.\,Affolder {\it et al.}, Nucl.\ Instrum.\ Meth.\ A {\bf 526}, 249 (2004).
                                                     
\bibitem{xft}
E.J.\,Thomson {\it et al.}, IEEE Trans. Nucl. Sci. {\bf 49}, 1063 (2002).
 
\bibitem{svt}
W.\,Ashmanskas {\it et al.}, 
Nucl.\ Instrum.\ Meth.\ A {\bf 518}, 532 (2004).

\bibitem{Punzi} 
G.\, Punzi,
eConf {\bf C030908}, MODT002 (2003);
arXiv:physics/0308063.


\bibitem{cdfsim}
E.\,Gerchtein and M.\,Paulini,
eConf {\bf C0303241}, TUMT005 (2003);
arXiv:physics/0306031.

\bibitem{MNR}
M.\,Mangano, P.\,Nason and G.\, Ridolfi, Nucl.\ Phys.\ B {\bf 373}, 295 (1992).



\bibitem{CDF-psiphiBR} F.\,Abe {\it et al} [CDF Collaboration],
 Phys.\ Rev.\ D {\bf 54}, 6596 (1996).

\bibitem{CDF-psiphiPol}  D.\,Acosta {\it et al.}  [CDF Collaboration],
 Phys.\ Rev.\ Lett.\  {\bf 94}, 101803 (2005).

\end{thebibliography}
\end{document}